\documentclass[a4paper]{article}

\usepackage{psfig}

\newcommand{\dfrac}[2]{\frac{#1}{#2}}

\title{Pattern formation by competition: a biological example}
\author{
{\Large M.~Bezzi}\\
SISSA - Programme in Neuroscience\\
Via Beirut 2-4 I-34103, Trieste, Italy\\
bezzi@sissa.it\\ \\  {\Large A.~Ciliberto }
and {\Large A.~Mengoni}\\
Dipartimento di Biologia Animale e Genetica dell' Universit\`a di Firenze,\\
 V. Romana 17, 50125 Firenze, Italy\\ }

\begin{document}
\maketitle

\begin{abstract}

We present a simple model based on a reaction-diffusion equation
to explain pattern formation in a multicellular bacterium 
({\sl Streptomyces}). We assume competition for resources as the basic
mechanism that leads to pattern formation; in particular we are able to
reproduce the spatial pattern formed by bacterial 
aerial mycelium in case of growth in minimal (low resources) and maximal
(large resources) culture media.  
\end{abstract}

\section{Introduction}

Bacteria are unicellular organisms
generally studied as isolated units,
however they are interactive organisms able
to perform collective behaviour, 
and a clear
marker of the presence of a multicellular organization level
is the formation of growth patterns~\cite{sha88,sha91}.
Particularly it has been pointed out that unfavorable conditions 
may lead bacteria to a cooperative behavior, as a means to
react to the environmental constraints~\cite{ben94}.

Many studies about the multicellular level of organization of
bacteria have
been proposed  and
pattern formation during colonies growth
has  been observed in Cyanobacteria~\cite{sha88}, in Bacillus
subtilis~\cite{sha88,mats93,ben97}, in Escherichia coli~\cite{sha88,ben95},
Proteus mirabilis~\cite{sha88,sha91} and others.
Some of these patterns have been studied by mathematical
models~\cite{ben94,mats93,ben97,ben95,mats97}, 
that explain the macroscopic patterns through
the microscopic observations.

There is a group of bacteria that differs from those
cited above because their normal morphological
organization is clearly multicellular: Actinomycetes,
and 
{\sl Streptomyces} is a genus of this group. 
{\sl Streptomycetes} are gram-positive bacteria that grow as mycelial
filaments in the soil, whose
mature colonies may contain two types of mycelia,
the substrate, or vegetative, mycelium and the aerial mycelium, 
that have different biological roles~\cite{Cha84}.
Vegetative mycelium absorbs the nutrients, and
is composed of a dense and complex network of hyphae usually 
embedded in the soil.
Once the cell culture becomes nutrient-limited, the aerial mycelium
develops from the surface of the vegetative mycelium.
The role of this type of mycelium is 
mainly reproductive, indeed the aerial mycelium develops the spores
and put them in a favorable position
to be dispersed~\cite{Cha84,Mig}.

\begin{figure}[f]
\centerline{
\psfig{figure=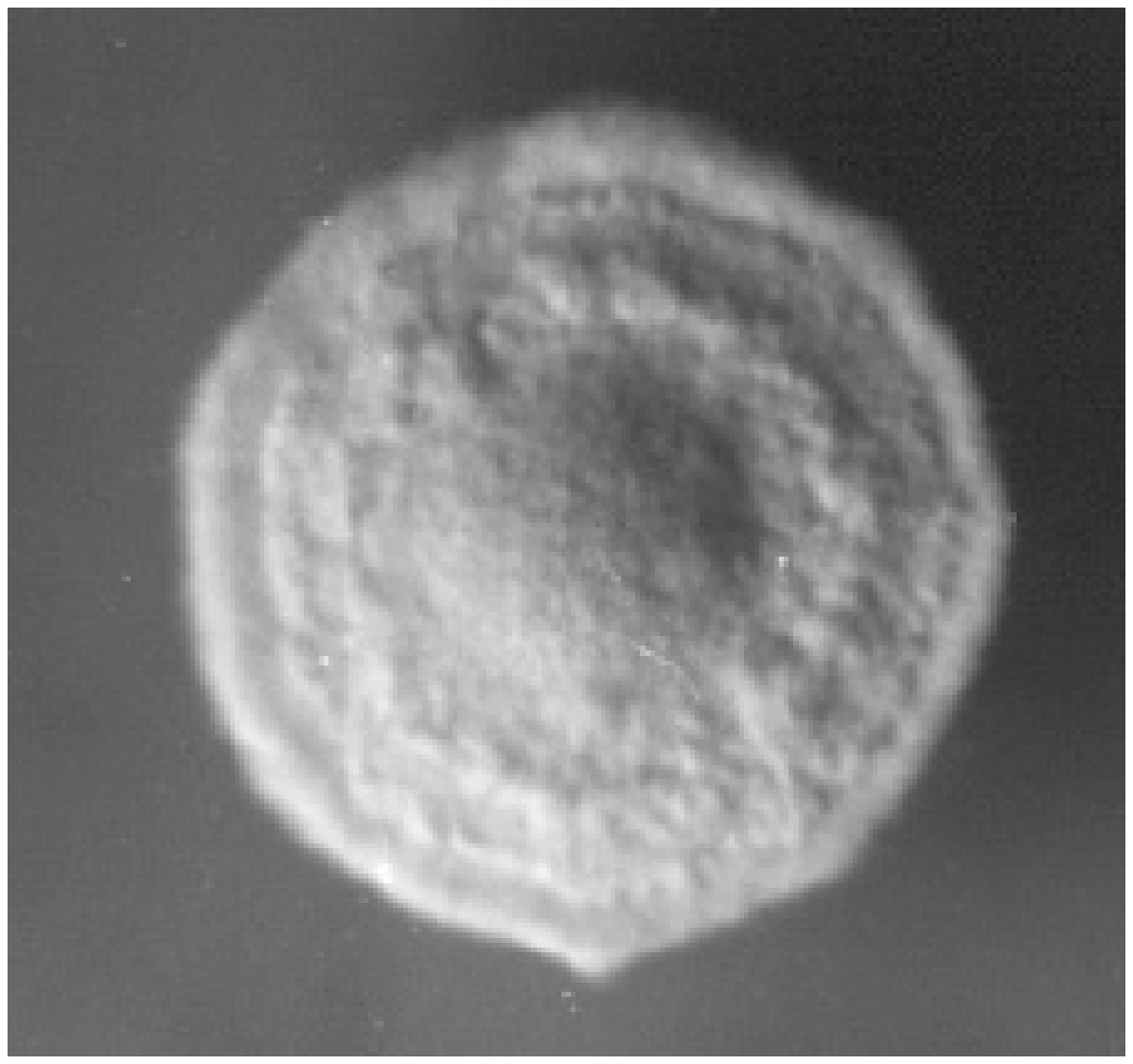,angle=270,width=8cm}
}
\caption{Pattern formed by {\sl Streptomyces} growing in maximal culture
  media. See details in the text.}
\label{exp:max}
\end{figure}

\begin{figure}[f]
\centerline{
\psfig{figure=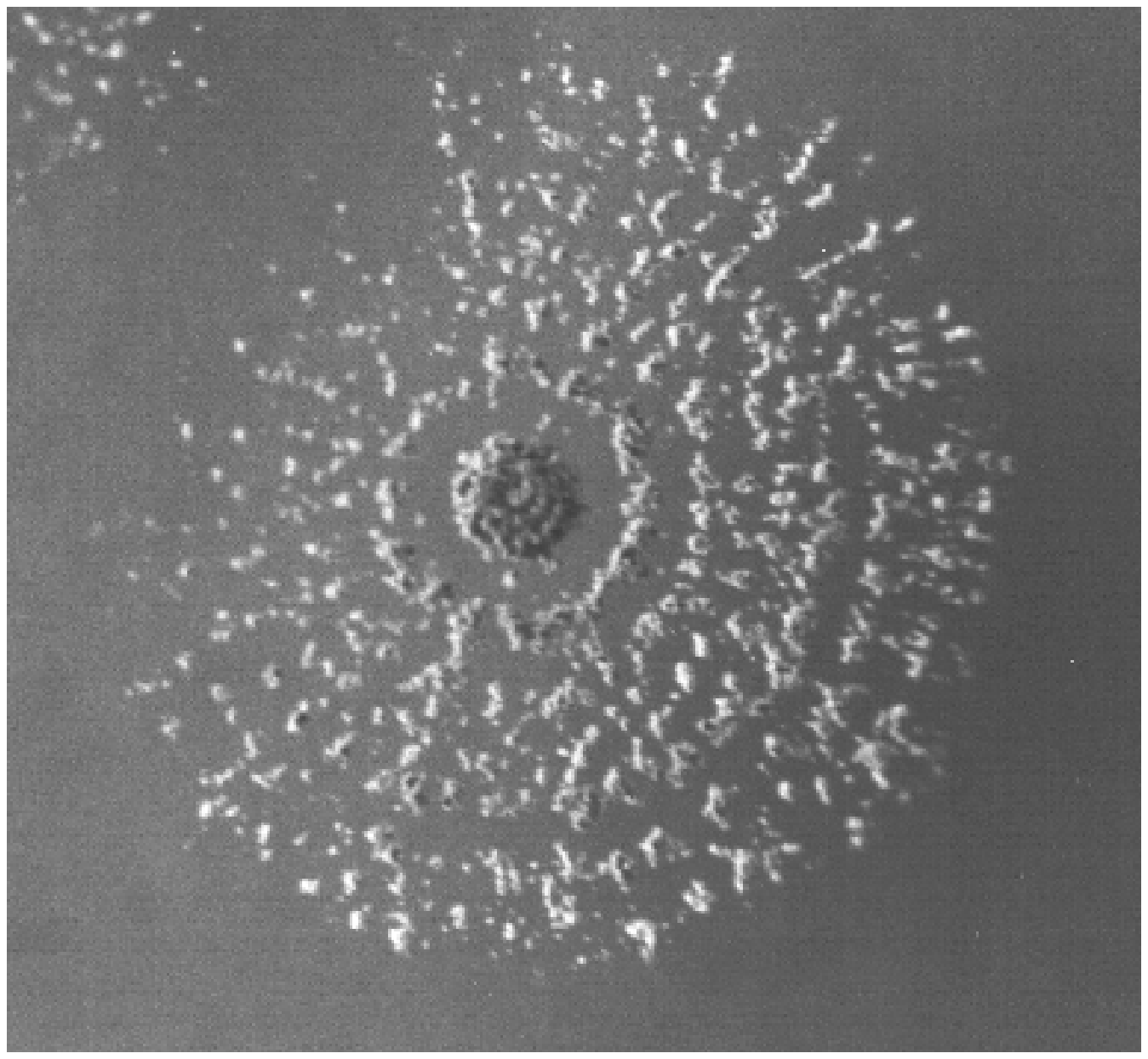,angle=270,width=8cm}
}
\caption{Pattern formed by {\sl Streptomyces} growing in minimal culture media.
 See details in the text.}
\label{exp:min}
\end{figure}

In our laboratory we have isolated a bacterial strain, 
identified with morphological
criteria as belonging to {\sl Streptomyces}. This strain is
interesting  because its growth pattern differs
on maximal and minimal culture media.
On maximal culture medium (LB, Luria Broth)~\cite{Sambrook1989}, 
after $3-4$ days of growth at $30^{\circ}C$, the strain
 shows a typical
bacterial growth with formation of the rounded colony characteristic of
most of the bacterial strains (Fig.~\ref{exp:max})~\cite{Cha84}.
On minimal culture medium (Fahreus)~\cite{Vincent1970} growth proceeds 
more slowly than in maximal media and a concentric
rings pattern of aerial mycelium sets up (Fig.~\ref{exp:min}).
The rings are centered on the first cell that sets up the
colony - we call it the founder - where usually the aerial mycelium
develops as well. The number of rings increases with time till $7-8$
after $20$ days of growth at $30^{\circ}C$. In both cases agar concentration was 
$1.5 \%$.

The presence of concentric rings patterns is a quite common feature in bacterial and
fungi colonies
~\cite{Adler66}; many models can originate such patterns~\cite{mats98},
a possible explanation was proposed in~\cite{Tsimiring95}, where is suggested that
the interplay of front propagation and Turing instability can
lead to concentric ring and spot patterns.  
A different approach based on competition for resources has been
recently proposed~\cite{BagnoliECAL,BagnoliPRL} to study species formation as
pattern formation in the genotypic space. We consider a similar mechanism to
investigate the spatial
pattern formations observed in our laboratory in a 
{\sl Streptomyces} colony.

\section{The model}

\subsection{Biological constraints}

Before introducing the mathematical model we have to go through some of the
biological features of the system.
Aerial mycelia are connected through 
the vegetative hypae network. 
This network has a peculiar structure in the {\sl Streptomyces}
isolated in our laboratory, indeed 
we observe that the growing boundary of the substrate mycelium
is made by many hyphae extending radially from the founder so that,
in this area, the substrate mycelium has a  radial polarity,
also if the hyphae have many branching segments.

Substrate mycelium has the biological objective
to find nutrients to give rise to spores,
therefore we expect that on minimal media a strong competition
arises for the energetic resources between neighbor substrate mycelia,
whereas in
maximal media, where there are sufficient nutrients, the
competition is weaker.

If the cells are connected mainly along the radial
direction, then competition will be stronger
along this direction than along the tangential one.
In other words, in the growing edge of the colony,
the competition is not isotropic
but, following the vegetative mycelium morphology, it will be stronger
among cells belonging to
neighboring circumferences (radial direction)
than among cells of the same (tangential direction), and we
will keep track of these aspects in the model.
Although the radial polarity is lost inside the colony,
the asymptotic distribution of aerial mycelium is strongly
affected by the initial spots derived by the growing boundary 
of the vegetative mycelium.

Finally another important feature of the biological system is the presence of a founder.
The founder behaves as every other aerial mycelium - it competes with the 
other cell -, moreover it is the center of every circle.
That means that every hypha originates from the founder: it is the source of the
vegetative hyphae, and as the colony grows the ring near the founder become
increasingly densely packed. 
Moreover during the enlargement of the colony no new center sets up
and therefore   
substrate mycelium density is highest near
the founder and decreases radially away from it. 

To summarize, in our model we make the following assumptions
based on the previous considerations.

\begin{itemize}
\item{There is competition among every aerial mycelium for some substances
that we assume for sake of simplicity uniformly distributed over the culture.}
\item{We consider only the aerial mycelium: we do not 
introduce explicitly the substrate mycelium but we take in account it assuming
that}
\begin{itemize}
\item[a)]{The competition  is stronger along the radial direction than along the
tangential one.}
\item[b)]{The probability for the aerial mycelium to appear is higher near the
founder}
\end{itemize}

\end{itemize}

Assuming this framework
we show that a concentric rings pattern
may be explained as a consequence of strong competition,
and a rounded pattern of weak competition.
From the biological point of view this
result implies that the formation
of concentric rings patterns is a mean that {\sl Streptomyces} adopts to
control growth.

\subsection{The mathematical model}

In the following we propose a mathematical model to reproduce the 
aerial mycelium growth patterns 
described in the Introduction.
This model is derived from a similar model introduced, in a different framework, 
(species formation in genotypic space) in~\cite{BagnoliECAL,BagnoliPRL}.

Let us consider a two-dimensional spatial lattice, that represents the Petri dish.
Each point ${\bf x}$ is identified by two coordinates ${\bf x}=(x_1, x_2)$, we study 
the temporal evolution of  the normalized probability $p({\bf x},t)$
to have an aerial mycelium in 
${\bf x}$ position at time $t$. The evolution equation for $p({\bf x},t)$, is in the form: 

\begin{equation}
p({\bf x},t+1)=A({\bf x},p({\bf x},t))p({\bf x},t)  \label{evol}
\end{equation}

where $A({\bf x},p({\bf x},t))$ is the probability of formation of a new aerial mycelium 
in position  ${\bf x}$  
and we suppose it can depend also on the distribution 
$p({\bf x},t)$. According to the hypothesis described above, it  is the product of two
independent terms: 

\[
A({\bf x},p({\bf x},t))=\frac{A_1({\bf x})A_2({\bf x},p({\bf x},t))}{\overline{A}} 
\]

where $A_1({\bf x})$ is the so-called static fitness, and represents the
probability of growth of an aerial mycelium 
in presence of an infinite amount of resources (no competition).
The founder is the source of every hypha, so we expect it will be
a decreasing function of the distance $|x|$ from the founder,
with $|x|=\sqrt{x_1^2+x_2^2}$,
assuming the founder occupies $(0,0)$ position.

The second term $A_2({\bf x},p({\bf x},t))$ is the competition term,
and in general it depends on the whole spatial
distribution $p({\bf x},t)$, 
moreover we suppose that two aerial  micelia  compete
as stronger as close they are. 

$\overline{A}$ is the average fitness and
it is necessary to have $p({\bf x},t+1)$ normalized. It is defined as
following
:

\[
\overline{A}(t)=\int_{\bf x} A({\bf x},p({\bf x},t))d{\bf x} 
\]

Both terms  are positive, therefore can be written in the exponential form

\[
A_1({\bf x})A_2({\bf x},p({\bf x},t))=\exp \left( H_1({\bf x})-J\int_{\bf y}
K(d({\bf x},{\bf y}))p({\bf y},t)dy \right) 
\]

where $J$ is the intensity of competition (it will be large in presence of
strong competition, i.e. low resource level) and $K(d({\bf x},{\bf y}))$ is a
decreasing function of the distance between two mycelia $d({\bf x},{\bf y})$.

We also allow $p({\bf x},t)$  to diffuse to the nearest neighbors with
diffusing coefficient $\mu$\footnote{The presence of diffusion is necessary to
allow the bacteria to populate the whole lattice}.

Finally we get:

\begin{equation}
p({\bf x},t+1)=\frac{\exp \left( H_1({\bf x})-J\int_{\bf y}K(d({\bf x},{\bf y}))
p({\bf y},t)d{\bf y}\right) }{\overline{%
A}(t)}p({\bf x},t)+\mu \nabla ^2p({\bf x},t) 
\label{evolution}
\end{equation}

According to the assumptions stated in Section 2.1, we now introduce 
the particular forms for $H_1({\bf x})$ and $K(d)$.
$H_1({\bf x})$
depends on the distance from the founder $H_1({\bf x})=H_1(|x|)$, and the
competition kernel $K(d)$, depending on the distance $d$ between 
mycelia.
As mentioned above, we expected the probability of growth 
for the aerial mycelium to be higher near the founder, therefore $H_1(|x|)$ has
to be a decreasing function of $|x|$. For the sake of simplicity we have chosen 
a single maximum, ``almost linear'' function,
\begin{equation}
H_1(|x|) =h_0 + b\left(1-\dfrac{|x|}{r} - \dfrac{1}{1+|x|/r}\right) \label{H_1} \\
\end{equation}

that has a quadratic maximum in ${\bf x}=(0,0)$ (founder), in fact close to 
${\bf x}=(0,0)$ we have
$h(|x|) \simeq h_0-b |x|^2/r^2$ and for $|x|\rightarrow \infty$,
is linear $h(|x|) \simeq h_0+b(1-|x|/r)$. $b$ and $r$ control the intensity of the
static fitness.

The competition kernel $K(d)$ has to be a steep decreasing function of $d$; 
we expect to have a finite range of competition, i.e. two mycelia at distance 
$d> R$ do not compete (or compete very weakly). 
A possible choice is:
 
\begin{equation}
K(d) =\exp\left(-\dfrac{1}{4} \left|\dfrac{d}{R}\right|^4\right)
\label{kernel}
\end{equation}

We have also chosen the form for the kernel (\ref{kernel}) and static fitness
(\ref{H_1}) because it is possible to derive
some analytical results~\cite{BagnoliPRL} that assure us the existence of a
non-trivial spatial distribution for exponential kernel with exponent greater
than $2$; $R$ is the range of competition.
All the numerical and analytical results described in this paper are
obtained using (\ref{H_1},~\ref{kernel}), but we have also tested similar potential
obtaining the same qualitative results.

Computing numerically from Eq.~(\ref{evolution})
the asymptotic probability distribution 
$p({\bf x}) \equiv p({\bf x},t)_{t \rightarrow \infty}$,
we get, for different values of the
parameters, two types of spatial patterns. 
In particular numerical and analytical studies (see
 Ref.~\cite{BagnoliPRL})
show that the crucial parameter is $G=\left( J/R\right) /\left( b/r\right) $,
i.e. the ratio between the intensity of competition and the intensity of the
static fitness.

\begin{figure}[f]
\centerline{
\psfig{figure=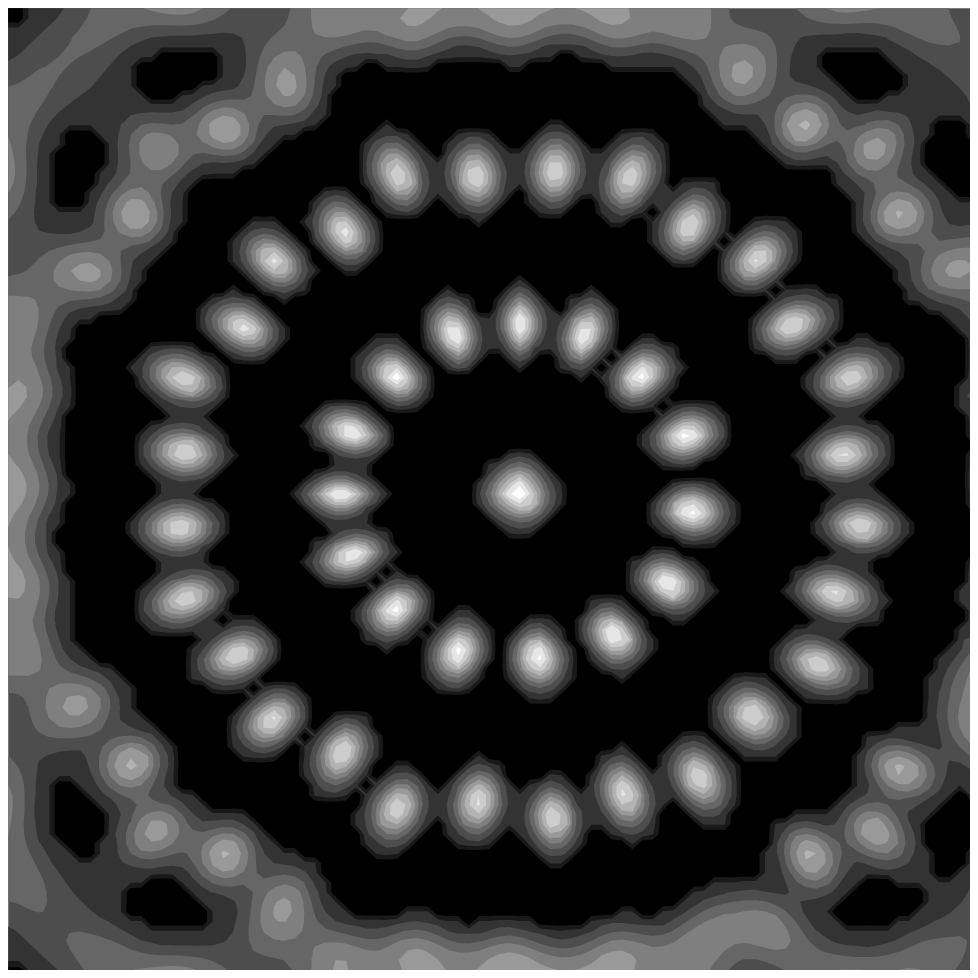,angle=270,width=8cm}
\hspace{.5cm}
\psfig{figure=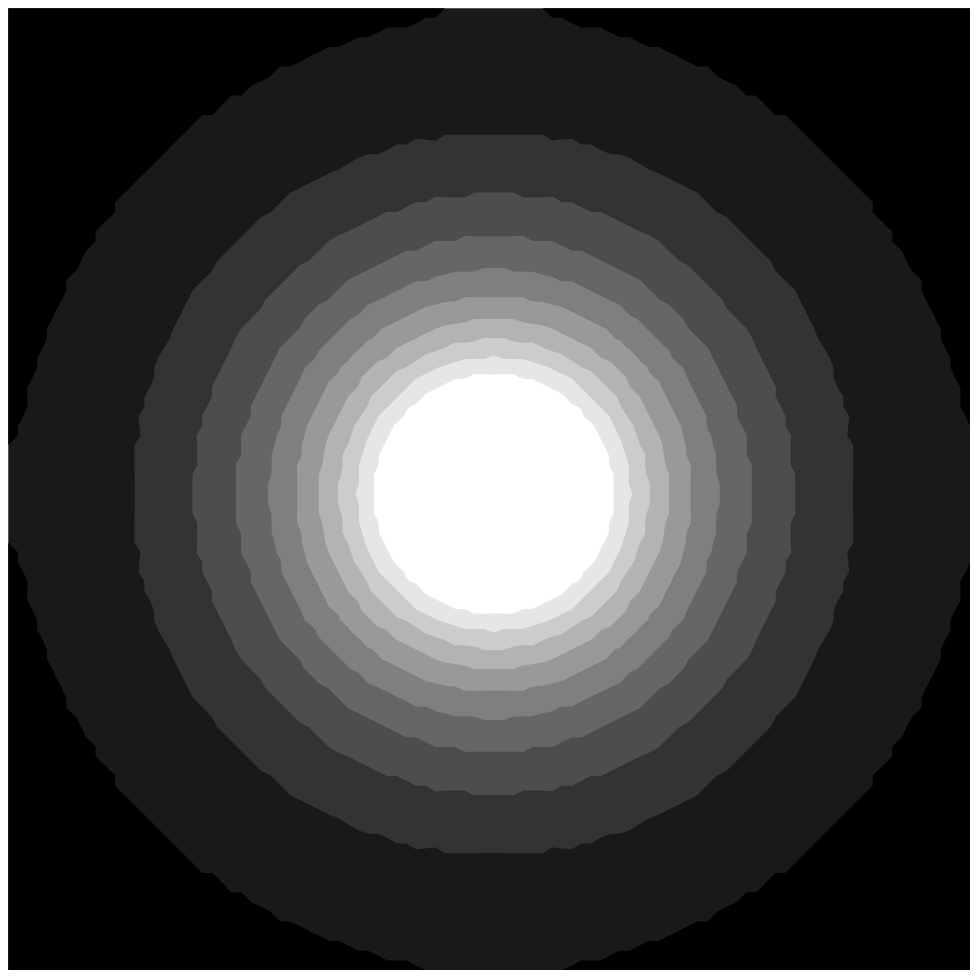,angle=270,width=8cm}
}
\caption{Asymptotic distribution $p({\bf x})$  for different 
values of parameters, plotted in inverse gray-scale, i.e. black $p({\bf x})=0$, 
white $p({\bf x})=1$. Left (large resources): $\mu=0.015$, $h_0=0$, $b=0.1$, 
$r=2$, $J=0.1$ and $R=4$. Right (low resources): 
 $\mu=0.015$, $h_0=0$, $b=0.05$, 
$r=2$, $J=56.0$ and $R=9$.}
\label{sim}
\end{figure}

For small values of $G$, that is the competition is rather weak or in other
words we have a maximal medium, we get a single peak
gaussian-like distribution centered on the founder  
(similar to the one showed on the left in Fig.~\ref{sim}~(left) with
$G= 0.5$). 

\begin{figure}[f]
\centerline{
\psfig{figure=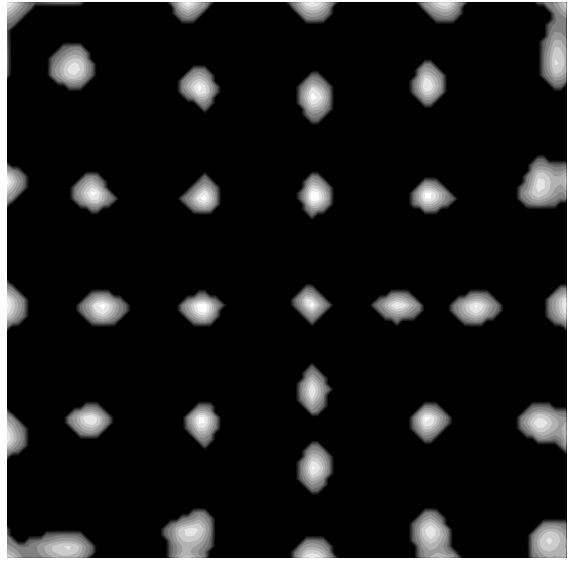,angle=270,width=8cm}
}
\caption{Asymptotic distribution $p({\bf x})$ for isotropic competition
($\alpha=1$) plotted in inverse gray-scale, i.e. black $p({\bf x})=0$, white
$p({\bf x})=1$, in  low resource case: 
 $\mu=0.015$, $h_0=0$, $b=0.05$, 
$r=2$, $J=56.0$ and $R=9$. The discretization of space (square lattice) 
for numerical solution of Eq.{~\ref{evolution})} is clearly evident.}
\label{alpha1}
\end{figure}

For
larger values of $G$ we get a multi-peaked distribution (see
Fig.~\ref{alpha1}, $G= 248.0$), where the central
peak (founder) is still present, but we get also some others peaks at an
approximate distance $R$, range of competition, between each other. This is
the expected pattern for an isotropic competition, in fact
the presence of equally distanced spots is due to the
competition term, that inhibits the growth of any
aerial mycelium around another one. 

To obtain spatial patterns similar to the concentric rings 
observed in our experiments, some feature of the  peculiar
spatial structure of {\sl Streptomyces} has to be added. As stated before,
we hypothesize that due to
the presence of the substrate mycelium morphology the
competition is much stronger in the radial direction (along the hyphae) than in
the tangential direction.

\begin{figure}[f]
\centerline{
\psfig{figure=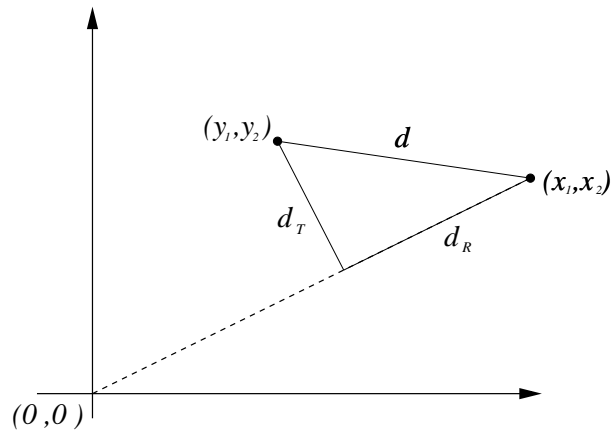,angle=270,width=8cm}
}
\caption{Decomposition of  the distance between any points ${\bf x}$ and 
${\bf y}$ in a radial $d_R({\bf x},{\bf y})$ and tangential part
 $d_T({\bf x},{\bf y})^2$ with respect to a circle centered in the founder
 placed in $(0,0)$.}
\label{Fig:distance}
\end{figure}

Therefore we decompose the distance between any points ${\bf x}$ and 
${\bf y}$ in a radial $d_R({\bf x},{\bf y})$
 and tangential part $d_T({\bf x},{\bf y})^2$ (see Fig.~\ref{Fig:distance})

\begin{equation}
d({\bf x},{\bf y})^2=  d_R({\bf x},{\bf y})^2 +\alpha d_T({\bf x},{\bf y})^2 
\label{newdist}
\end{equation}

where $\alpha$ is a parameter that allows to change the metric of our space.

For $\alpha > 1$ 
the relative weight of tangential distance is larger than
one due to the lack of cell communications along this direction,
the competition is mainly radial along the hyphae because the mycelia do not
compete if they are not directly connected by an hypha.
For $\alpha = 1$ we get the usual euclidean distance.

Using the distance (\ref{newdist}) in Eq.(\ref{evolution}) with  $\alpha > 1$
and strong competition
 we are able to obtain a set of rings
composed by equally spaced spots 
at fixed distances from the founder 
(see Fig.~{\ref{sim}}~(right) for $\alpha=6$), while in presence of large resource 
we still have a single peaked distribution (Fig.~{\ref{sim}}~(left)).
For larger  values of $\alpha$ the rings become continuous, while for low values,
$\alpha \rightarrow 1$, 
the multi-peaked structure of $p({\bf x})$ appears.

These results are in agreement with those presented in 
 Ref.~\cite{BagnoliPRL}, where an one-dimensional system is considered. 
 In this case the genotypic space plays the role of the real space, 
 and using  and a gaussian kernel  
\[
K(d) =\exp\left(-\dfrac{1}{2} \left|\dfrac{d}{R}\right|^2\right)
\]
 is possible to derive analytically the value of transition $G_c$
 between the two regimes (single
 peaked and multi-peaked distribution). It is, for $\mu \rightarrow 0$ (slow
 diffusion) and $\frac{r}{R} \rightarrow 0$ (static fitness almost flat)
\[
        G_c(\frac{r}{R}) \simeq G_c(0) - \frac{r}{R} 
\]
with $G_c(0) = 2.216\dots$. 
Thus for $G>G_c(\frac{r}{R})$ we have a multi-peaked distribution, while 
for $G<G_c(\frac{r}{R})$ only the fittest one survives (single-peaked
distribution).

\section{Discussion and conclusions}

We isolated a strain of {\sl Streptomyces} that has a dual pattern of
growth concerning the aerial mycelium: it gives rise to concentric rings
centered on the founder cell, or to the classic circular bacterial colony. 
The medium is discriminant: in minimal media the first type of pattern arises, in
maximal media the second one. 

The substrate mycelium follows a different
pattern:  optical microscopy observations revealed that every hypha
originates from the primordial central colony (the founder). Moreover the
growth of the substrate mycelium growing edge 
proceeds in radial direction from the founder. 

Using a simple mathematical model for the formation of aerial mycelium we 
are able to simulate both aerial mycelium spatial patterns.
 The parameter we modulate to obtain these two different patterns is the 
competition intensity. Indeed
the main assumption of the model is that there is competition among the 
hyphae of vegetative mycelia for the energetic sources necessary for 
the formation of the aerial mycelium. In a medium with
low nutrient concentration there is a strong competition for the aerial 
mycelium formation - and the model 
produces concentric rings patterns - instead in a maximal medium the
competition is weaker - and the model produces the classic circular 
bacterial colony.

The aerial mycelium is derived by the substrate mycelium, so
we derived the constraints of the model from the morphological 
observations concerning the substrate mycelium described in the
Introduction. The 
system has a radial geometry centered on the founder (the probability of 
formation of aerial mycelium is higher near the founder), and we assumed 
that the competition is affected by this feature. Indeed the 
competition is stronger along an hypha due to the cell-cell 
communication typical of the ``multicellular'' organization of {\sl 
Streptomyces}. This implies that the competition is stronger along the 
radial direction than along the tangential, at least in the outer boundary
of the colony. 

The growth pattern description above is referred to the presence of one single
primordial colony.
In presence of two or more colonies close one another we have
observed different patterns with additive and
negative interactions among the colonies. 
Our minimal model is not able to reproduce these behaviors, due to the
fact that in presence of many founders 
the simple
assumptions of radial growth centered on a single
founder is no more fulfilled.

\begin{figure}[f]
\centerline{
\psfig{figure=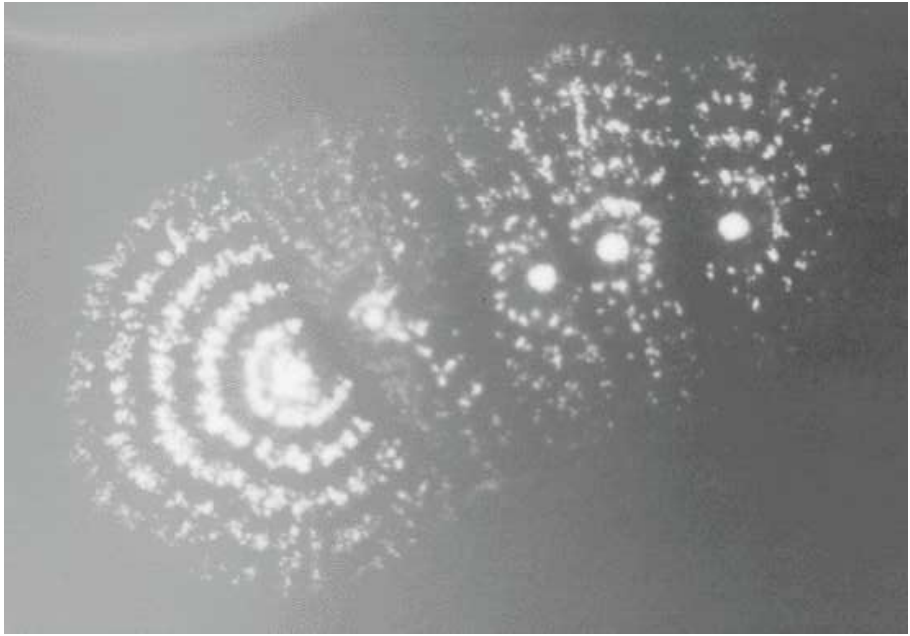,angle=270,width=12cm}
}
\caption{Interference pattern formed by {\sl Streptomyces} colonies growing in minimal culture
  media. }
\label{exp:int}
\end{figure}

In conclusion we have found some peculiar spatial patterns
for the aerial mycelium of {\sl Streptomyces}. We have proposed a simple
 mathematical  model to explain these patterns assuming competition along
 the hyphae as  the main ingredient that leads to pattern formation.
Our numerical results are able to reproduce spatial patterns obtained
 experimentally under different conditions (minimal and maximal medium), while
 to get more complex behavior (interference patterns, see Fig.~\ref{exp:int}) we expect 
more ``chemical'' species have to be added to our minimal model.

\section*{Acknowledgements}
We wish to thank F. Bagnoli, M. Buiatti, R. Livi and A. Torcini for  
fruitful discussions.
M.B. and A.C. thank the Dipartimento di Matematica Applicata ``G. Sansone'' for 
friendly hospitality.

\newpage

\end{document}